\documentclass[%
 reprint,
 amsmath,amssymb,
 aps,
]{revtex4-1}

\usepackage{graphicx}
\usepackage{epstopdf}
\usepackage{dcolumn}
\usepackage{bm}

\begin{document}

\preprint{APS/123-QED}

\title{A Quantum Phase Representation of Heisenberg Limits \\ and a Minimally
Resourced Quantum Phase Estimator}

\author{Scott Roger Shepard$^{1}$, Frederick Ira Moxley III$^{2}$, and Jonathan P.
Dowling$^{2,3}$}
\affiliation{
$^1$ Department of Electrical Engineering, College of Engineering \&
Science\\
Louisiana Tech University, Ruston, LA 71272, USA\\
$^2$ Hearne Institute for Theoretical Physics, Department of Physics
\& Astronomy\\
Louisiana State University, Baton Rouge, LA 70803, USA \\
$^3$ Beijing Computational Science Research Center, Beijing 100084,
China
}

\date{\today}

\begin{abstract}
Within the quantum phase representation we derive Heisenberg limits, in closed form, for N00N states and two other classes of states that can outperform these in terms of local performance metrics relevant for multiply-peaked distributions. One of these can also enhance the super-resolution factor beyond that of a N00N state of the same power, at the expense of diminished fringe visibility.  An accurate phase estimation algorithm, which can be applied to the minimally resourced apparatus of a standard interferometer, is shown to be resilient to the presence of additive white-Gaussian noise. 
\begin{description}
\item[PACS numbers]
42.50.St, 42.50.Dv, 42.50.Ex, 42.50.Lc
\end{description}
\end{abstract}

\maketitle


\section{Introduction}

Quantum phase estimation plays an important role in quantum
computing and quantum sensing applications. With regard to quantum
computing, phase estimation is often implemented via the inverse of the
Quantum Discrete Fourier Transform (QDFT). Phase estimation via QDFT is an
integral step in many quantum algorithms, including the assessment of
periodicity in Shor's algorithm, etc. [1, 2]. Quantum computing is
typically realized as quantum circuits implementing algorithms on individual
qubits [3], however, these can also be realized as quantum multiparticle
interferometers [4, 5] along the lines of Feynman's original ideas [6]. A
mapping between quantum circuits, interferometers, and spectrometers [7],
coupled with linear optics realizations [8], leads naturally to a quantum
sensing perspective that is realizable via beam splitters, phase shifters
and photodetectors.
\newline
\indent 
Therein, in addition to the acheivement of
phase measurement accuracies below the shot-noise limit [9], the quantum interference effects in electromagnetic fields have led
to super-resolving phase measurements [10, 11] (a.k.a. super-resolution), which also 
circumvents the Rayleigh diffraction limit in lithography [12] and imaging [13]. 
In essence, $N$
photons of a field at wavelength $\lambda $ are utilized to perform
``quantum sensing'' with an effective wavelength of $\lambda /N$, while
still utilizing sources, detectors and propagation properties associated
with an actual wavelength $\lambda $. In the standard methods, the
improvement factor $N$ has been limited to date [14, 15, 16]. This
limitation arose primarily because
in order to observe these higher-order fringes, the standard schemes relied on
coincidence detection methods. Thus, the measurement apparatus increases in
complexity with $N$. It was thought that coincidence
detection schemes of order $N$ were required since the super-resolving fringes vanish in the output of  a first-order interferometer. Fortunately, however, a method of extracting
this higher-order phase information from a standard  [17] interferometer has been found [18]. In this method (termed the
phase function fitting algorithm, PFFA) the apparatus complexity is
independent of the super-resolving improvement factor $N$. Herein, we examine the robustness of the PFFA to
additive white-Gaussian (e.g., thermal) noise --- after first gleaning
insights, and deriving N00N (and other state) limits within the quantum phase
[19, 20] representation.
\newline
\indent
The quantum phase representation that is complementary to the measurement of
the difference of photon numbers between two harmonic oscillators is useful
for visualizing and calculating the phase information associated with a
quantum state -- although its apparatus has yet to be realized. Herein
it is used to derive Heisenberg limits (without bounds or approximation)
for three classes of states; and to show how an entanglement of N00N
states with the vacuum state can surpass the N00N state in terms of two
local performance measures appropriate for multiple-peaked distributions.
Both measures, 
the HWHM and
the square-root of the bin-variance, 
scale as $\sim 1/N$ (where $N$ is the
average photon number) for all three classes of states --- which are
Heisenberg limited in that sense. They differ however in the coefficient
which multiplies $1/N$ and one of these  can go to zero, but
only at the expense of reduced fringe visibility. 
\newline
\indent
The simplest phase estimation apparatus that we know how to realize is a
standard quantum interferometer. Therein quantum resources are diminished by
exploiting the multiphoton interferences inherent within the probability
amplitudes of the quantum electromagnetic field itself.
The quantum theory of an interferometer is based on the  observation by Yurke et al. [21] that it is mathematically isomorphic  to rotating a quantum state by an unknown angle and estimating that angle from the projection of the rotated state onto the z-component angular momentum eigenkets. This stems from Schwinger's observation that the algebra of two uncoupled harmonic oscillators can reproduce the algebra of angular momentum [22]. 
 In terms of eigenvalues we have $m=(n_u - n_d)/2$ for the eigenvalue associated with $\hat{J}_z$ and $j=(n_u+n_d)/2$ for the eigenvalue  associated with $\hat{J}^2 \equiv \hat{J}^2_x + \hat{J}^2_y +\hat{J}^2_z$, where  $n_u $ and $n_d$  are the photon number eigenvalues of the two oscillators. 
The interferometer’s statistics are $P_{m} = |\textbf{$\Psi$}_{m}(\textbf{$\Phi$})|^{2}$  which is a probability distribution in discrete $m$ space, under relative interferometer-arm phase shift \textbf{$\Phi$} and the underlying wavefunctions are $\Psi_{m}(\textbf{$\Phi$})\equiv\langle{m}|\hat{D}_{x}(\textbf{$\Phi$})|\psi\rangle$ where $\hat{D}_{x}(\textbf{$\Phi$})$  is the analogous rotation about the x-axis by $\textbf{$\Phi$}$, the unknown signal we wish to estimate.  
The general theory of quantum angle measurement (complementary to the measurement of a single component of angular momentum) is described in [19, 20] but for states of unique $j$ for all $m$ the general theory reduces down to what one would naturally expect for complementary quantities: a Fourier transform between wavefunctions. 

\section{Quantum Phase Representation of Three Classes of States}

Although its apparatus has yet to be realized, the phase (or angle) representation is
useful for visualizing and calculating the phase information associated with
a quantum state. Herein we use it to derive Heisenberg limits
(without bounds or approximation) for three classes of states --- two of
which can be written in the following form (times a normalization constant) 
\begin{align*}
&|\psi\rangle = r_2 (|{2  j_{\rm{max}}\rangle}_u |0\rangle_d + |0\rangle_u |{2 j_{\rm{ max}}\rangle}_d) +  \\
& r_1 (| j_{\rm{ max}}\rangle_u |0\rangle_d + |0\rangle_u | j_{\rm{ max}}\rangle_d) +|0\rangle_u |0\rangle_d. \tag{1}
\end{align*}
The first of the three classes considered are the N00N states; the second class (termed sub-states) are
an entanglement of a N00N state of $j = j_{\rm{ max}}/2$, with an equally likely
superposition of a N00N state of $j = j_{\rm{ max}}$ and the vacuum state ($r_{2}=1/%
\sqrt{2}$); and the third class (termed N00N-vac states) are comprised of an entanglement of a N00N state with
the vacuum state ($r_{1}=0$). In all cases the cost function is the expected
number of photons used, $N=2\left\langle j\right\rangle $, and our metrics
are local performance measures appropriate for multiple-peaked distributions
(half-width-half-max, and bin-variance) as well as fringe visibility and
other aspects of the probability distribution function (PDF) of the quantum
phase measurement $P\left( \phi \right)$, including the number of peaks and their height.

\begin{figure}
\centering 
\includegraphics[scale=.85]{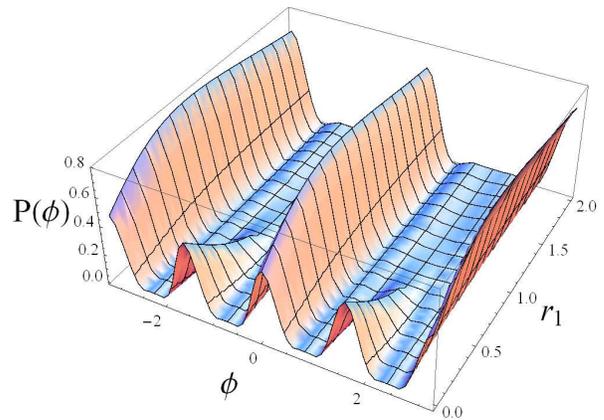}
\caption{(color online). An illustration of how the free parameter $r_1$ can affect the probability density function for a sub-state.}
\end{figure}

The Fourier transform of the number-ket expansion coefficients of the N00N
state readily yields $P({\phi}) = (1/\pi) {\rm{cos}}^{2}(j_{\rm{max}}{\phi}) $ which has $N=2j_{\rm{max}}$
peaks (of height $1/\pi $) separated by perfect nulls (hence the fringe
visibility $V \equiv (\max -\min )/(\max +\min )$ is always unity for N00N states in the phase representation, independent of $j_{\rm{ max}}$) and
we observe the super-resolving aspect of obtaining $N$ identical
peaks or fringes, in contrast to a single-peaked PDF which would arise in
the case of a coherent state. As a consequence of the Fourier transform the
periodicity (here the number of peaks, which is also the number of identical
bins) is set by the minimal separation of values of $m$ for which we
have non-zero number-ket expansion coefficients (one for the coherent state, 
$2j_{\rm{ max}}$ for the N00N state). The variance on a 2$\pi $ interval of this
multiple-peaked PDF is clearly not a useful performance measure, so we
consider bin-variance: defined to be the variance of the PDF over one of
these identical bins, renormalized to the bin width (for N00N states
that is a 2$\pi /N$ interval) and centered on that bin (to avoid branch-cut
effects). Inherent to the utility of this metric (and to the use of
super-resolution in general) is the assumption that we can correctly
assess the bin to which any particular estimate corresponds --- otherwise we make
a bin error (the probability of which will be influenced by fringe
visibility and how one tracks a dynamically varying unknown phase). We
similarly consider the local (defined on one bin) half-width at half-max,
HWHM, and obtain the N00N state results 
\begin{equation}
\rm{HWHM} = \frac{\pi }{2N}, \ \ \ \
\text{bin-variance}=\frac{\pi ^{2}/3-2}{N^{2}}.  \tag{2}
\end{equation}
The square-root of the bin-variance and HWHM both scale as $1/N$ in terms of our cost function. Indeed all three
classes of states considered herein follow a $1/N$ scaling but the
difference in the coefficient multiplying $1/N$ varies widely and can even
go to zero at the cost of diminished fringe visibility. Other differences,
gleaned from the PDFs, will impact upon the probability of bin error in a
practical system; but the fact that the other two classes of states can
yield coefficients better than those in equation (2) indicates that state
optimization remains a field open for investigation (e.g., the N00N state
is not optimal, even in the above two metrics).

\begin{figure}
\includegraphics[scale=.68]{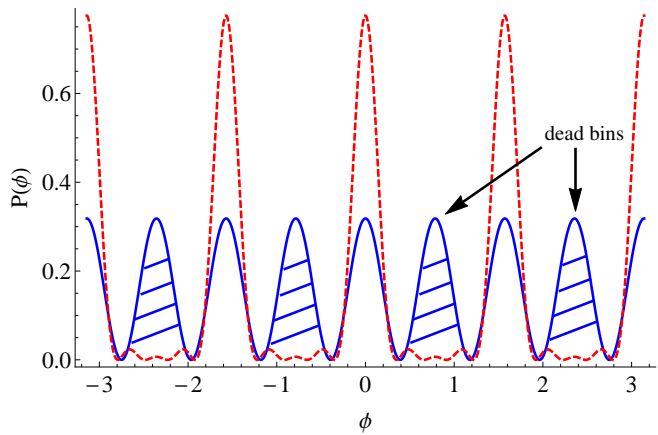}
\caption{(color online).   N00N (solid) and sub-state PDFs for N = 8. Sub-states have: (HWHM) (N) = $2\cos ^{-1}[(4r_1 -4\sqrt{6+4r_1 +{r_1}^2})/(8\sqrt{2})] \ \rm{and}$  \ ( \text{bin-variance})  $(\rm{N}^2) = [128(27+13\sqrt{2})r_1 +144(3+\sqrt{2})\pi ^2 r_1 + 36\pi^3 (1+{r_1}^2 )-27\pi (8\sqrt{2}-1+8{r_1}^2 )]/[36(4(3+\sqrt{2})r_1 + 3\pi (1+{r_1}^2 ))]$}
\end{figure}

Within the second class of states  we  interfere a 
N00N state (of $j = j_{\rm{ max}}$) with a sub-harmonic (hence the term sub-states) which will be a N00N state (of $%
j = j_{\rm{ max}}/2$) to, in effect, delete alternate bins and sharpen the PDF within
the remaining bins. Within this class we also interfere the entangled N00N states with the vacuum state and take $r_{2} = 1/\sqrt{2}$ so that the vacuum and
the larger ($j = j_{\rm{ max}}$) N00N state are equally likely, thus $\left\langle
j\right\rangle = j_{\rm{ max}}/2$ independent of $r_{1}$. In Figure 1 we illustrate
how the free parameter $r_{1}$ can affect the PDF (seen here for $ j_{\rm{ max}} = 2$, as 
$r_{1}$ ranges from 0 to 2).
When $r_{1}=0$ the state also belongs to the third class of states
(described later) and when $r_{1}\thickapprox 1$ we observe the sub-harmonic
effect, as demonstrated in Figure 2 in which we compare the PDF of an $N=8$
sub-state ($r_{1}=1$, $j_{\rm{ max}}=8$) to that of a N00N state of $N=8$ ($
j_{\rm{ max}}=4$). 
The coefficient multiplying $1/N$ in the HWHM for the
sub-states is given in the Figure 2 caption and it
monotonically increases in $r_{1}$ from $\pi /3$ at $r_{1}=0$; to the
N00N state limit of $\pi /2$ as $r_{1}\rightarrow \infty $ (as it must since in that
limit the sub-state approaches a N00N state). In addition to increased
sharpness reflected in the HWHM, examination of Figures 1 and 2 suggests that for $%
r_{1}$ near $1$ (where alternate bins are suppressed) the enhanced
peaks of the sub-states could also permit more rapid acquisition of useful
phase estimates during the collection of the histograms which evolve into
these PDFs (and higher peaks can be useful for tracking a dynamic unknown phase).
Moreover we might adopt a protocol of ignoring any estimates which belong to
the suppressed (a.k.a. dead) bins. Dropping the dead bins in this way is
not unduly complicated since in all these schemes we must choose the right
bin anyway to avoid a bin error, but it does come at the cost of dropping
some estimates, which occurs with probability $P(\rm{drop})=1/6[
3-4(3+\sqrt{2})r_{1}/\pi (1+r_{1}^{2})]$. Since $P(\rm{drop})$ achieves its
minimal value of about $0.0316374$ at $r_{1}=1$ this is not an unreasonable
protocol to adopt --- in which case the appropriate bin width is that of the
kept bin (e.g., in Figure 2 that would be $2\pi /8$, rather than $2\pi /4$).
The bin variance on the kept bin has a coefficient multiplying ($N^{-2}$)
equal to the lengthy expression in the Figure 2 caption, 
which increases monotonically from about $0.711441$ at $r_{1}=0$; to (again)
the N00N state limit as $r_{1}\rightarrow \infty $, which is $\pi
^{2}/3-2\thickapprox 1.28987$; and notably takes on the value of about $%
0.869983$ at $r_{1}=1$. Of course as $r_{1}\rightarrow 0$ we would not
employ the protocol of dropping alternate bins since in that limit these
have the same PDF as the kept bins; and $P(\rm{drop})$ would go to $1/2$
indicating that we'd be (pointlessly) dropping $1/2$ of our equally useful
estimates.
The third class of states considered herein have $r_{1}=0$, and $r_{2}$ is a
free parameter. These are entanglements of a N00N state (of $j=j_{\rm{ max}}$)
with the vacuum state, termed N00N-vac states. To clarify the physics
we let $r_{2}=1/\sqrt{2n}$ and initially consider $n$ to be an element of
the set of integers. For $n=1$ the N00N-vac state is equivalent to a
sub-state of $r_{1}=0$. The number of bins for this class of states is $%
j_{\rm{ max}}$ (the minimal distance in $m$ of the N00N state component is $%
2j_{\rm{ max}}$; but that distance is reduced to $j_{\rm{ max}}$ via inclusion of the
vacuum state) and there is no motivation for any bin dropping protocol. The
strategy for the N00N-vac states is to make $j_{\rm{ max}}$ large (to reap the
benefits of super-resolution) but at the same time reduce the
probability of {\it{actually}} being found in the N00N state component, by also
increasing $n$, since then we can simultaneously constrain $N\equiv
2\left\langle j\right\rangle$, which is $2\left\langle
j\right\rangle = 2j_{\rm{ max}}/(n+1)$ for these states. We find that the
probability amplitude of the vacuum state can strongly interfere with the
probability amplitude of the N00N state component even when the
probability of actually being in that component is greatly diminished; and
that the only fundamental tradeoff in this scheme is a slowly diminished
fringe visibility as $V = (2\sqrt{2}\sqrt{n})/(2+n)$.
\begin{figure}
\centering 
\includegraphics[scale=.9]{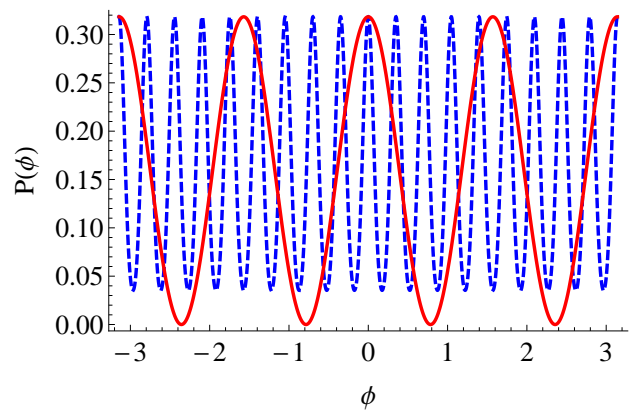}
\caption{(color online).   N00N (solid) and N00N-vac state (dashed)  PDFs  for $N = 4$ (and $n =8$). N00N-vac states have: (HWHM) (N) = $2/(n+1)\cos ^{-1}[1/2(-\sqrt{2}\sqrt{n}+\sqrt{2+2\sqrt{2}\sqrt{
n}+n})] \ \ \rm{and} $  (\text{bin-variance}) $(\rm{N}^2 ) = 2(3-24\sqrt{2}\sqrt{n}+2\pi^2 +2n\pi^2 )/[3(n+1)^3]$}
\end{figure}
For example, consider the ways in which we can make $\left\langle
j\right\rangle =2$ (i.e., an $N=4$ photon state). A N00N state with $%
j_{\rm{ max}}=2$ would produce $4$ bins; but since $1/4^{\rm{th}}$ of $8$ is also $2$
we could (for the same $N=4$ cost) produce twice as many bins with a N00N%
-vac state of $n=3$ and $j_{\rm{ max}}=8$, in which case the 
peaks as well as the range (i.e., the value of
$\rm{ max}$ - $\rm{ min}$) for the N00N-vac state are larger than those of the N00N state
(although $V<1$ since the minimum no longer goes to zero). In the next section we will
consider an algorithm which does not require the nulling of an
interferometer --- in such cases how close min comes to zero is of diminished importance. Moreover, we often are more concerned about the peaks of
these PDFs because their enhancement permits more rapid acquisition of
useful histograms (which is crucial while tracking a dynamic phase). It can
be shown that as we increase $n$ the peaks of the N00N-vac PDFs come down and become equal to those of the N00N states when $n=8$ (independent of $j_{\rm{ max}}$) as illustrated for $N=4$
in Figure 3. As we continue to increase $n$ the super-resolving enhancement
aspect (i.e., the number of fringes or bins at a fixed $N$) continues to
increase while the visibility diminishes, until at $n=2(17+12\sqrt{2})$ $%
\thickapprox 67.9411$ the $\rm{min} = \rm{max}/2$ (so that the visibility $V = 1/3$) and
beyond this value of $n$ the HWHM metric has no meaning. For $n$ less than
this value, the formula for HWHM has a coefficient (multiplying $N^{-1}$)
given in the Figure 3 caption, 
and is better
than $17.87$ times smaller than the N00N state limit at $n=67$. 
Similarly, the coefficient (multiplying $N^{-2}$) in the formula for the
bin-variance of the N00N-vac states:
\begin{equation}
\frac{2(3-24\sqrt{2}\sqrt{n}+2\pi ^{2}+2n\pi ^{2})}{3(n+1)^{3}},  \tag{3}
\end{equation}
is smaller than the N00N state limit (and is monotonically decreasing) for
all $n\geq 1$. Bin-variance maintains its meaning, even when $\rm{ min}>\rm{ max}/2$,
and its corresponding coefficient goes to zero as $n\rightarrow \infty $. 
With $V>1/3$ at $n=67$
the bin-variance coefficient is better than $569.9$ times smaller than the N00N state limit.

\section{The Phase Function Fitting Algorithm}

The phase function fitting algorithm [18] consists of the following protocol.
Under a given, but unknown, value of $\Phi $ one measures the
interferometer's statistics and retains the $2 j +1$ probabilities for each 
possible value of $m$: $P_{m}=\left|\Psi _{m} (\Phi) \right| ^{2}$.
To incorporate knowledge of the allowable quantum results one then calculates the
interferometer's statistics for some dummy variable $x$: $f_{m}(x)=\left|
\Psi _{m}(x)\right| ^{2}$  (for the four-photon N00N state of our simulations herein, those are $2j+1=5$ different
functions). One then performs an LMS (Least Mean Square error) fit of these $2j+1$
functions to the measured $2j+1$ numbers to perform an optimal estimation of
parameter $x$, thereby yielding our estimate of $\Phi $.

In so doing, our simulations yield  surprisingly good results: over
9 digits of phase accuracy, for a N00N state input of $j=2$, when $\Phi $
ranges over a bin-width, here an interval of $\pi /4$ [23]. 
At first this might seem to be either: an impossibly good result (over nine
digits of accuracy from only four photons would be orders of magnitude below the Heisenberg
limit of $1/4$); or a ridiculously trivial result (certainly similar
extractions of information might be made from the ideally collected
statistics of other measurements). There is no contradiction here, instead
it teaches that: 

\noindent {\bf 1) }{\it Local performance measures are important
characterizations of measurement statistics, but they do not uniquely
identify what can be achieved when further processing (an algorithm) is
applied; } and

\noindent{\bf 2)} {\it {most significantly for phase estimation, it specifically demonstrates that
the super-resolving information has not vanished from
the measurement performed in a quantum interferometer.}} 
These higher-order
fringes only vanished in the standard (first-order) way of extracting information from the
 interferometer statistics. Since the higher-order information is
uniquely extractable (over a non-trivial range) this information has not
been destroyed in the measurement process. This significantly opens the door
to the possibility of minimally resourced quantum interference sensors in
which the quantum resources are dramatically reduced via classical signal
processing that incorporates knowledge of the allowable quantum results. 

There is also a sense in
which an infinite number of photons have been used because we have extracted
these virtually error-free estimates from the calculated (equivalent to
ideally collected) quantum measurement statistics. In a laboratory
environment it would take time (hence many photons) for the collected
histograms to converge to the calculated PDFs. The unknown $\Phi $ cannot
vary appreciably during this time for these static limit results to hold.
This is the scenario within which most analyses are performed and it is how
a fringe pattern is often measured. 
Also $N = 4$ in these simulations and power (i.e., $N$, not energy) is the relevant constraint. It is power
that determines: the shot noise; the radiation pressure on a mirror; the
cost of the laser; etc., so $N$ is the relevant cost function. If we use
more energy $E \sim NT$ (even towards an infinite
number of photons) by collecting longer, that doesn't affect our cost and we
can do so provided we can still make $T$ small enough to track the dynamics.
To go beyond this static limit one would also need
to simulate the tracking of a dynamic signal, while acquiring and processing
the histograms. 

It is  significant to note that the phase information from such states
is ultimately extractible from the simplest phase estimator we know how to
build --- a standard interferometer. Other than what is required to create such 
states in the first place; there is no need to over burden the quantum
resources -- we can delegate some of the task to signal processing that is
armed with quantum knowledge. This is not a 
trivial result because it cannot be applied to arbitrary quantum measurements. If 
for example, our
apparatus consisted of a first-order interferometer (in which we subtract
and average the two photodetector currents) the desired phase information vanishes
entirely and no further processing can extract it. If instead we multiply
those two currents (second-order coincidence detection) or do a parity
measurement [24] then other limits are obtained. Note in passing that the
PFFA does not require any presently unachievable amount of number resolving
in the photodetectors. Indeed, in the four-photon ($j=2$) example presented
these need only discern between $0,1,2,3,$ or $4$ counts (photodetection events
within the   observation time $T$). 

\begin{figure}
\includegraphics[scale=.56]{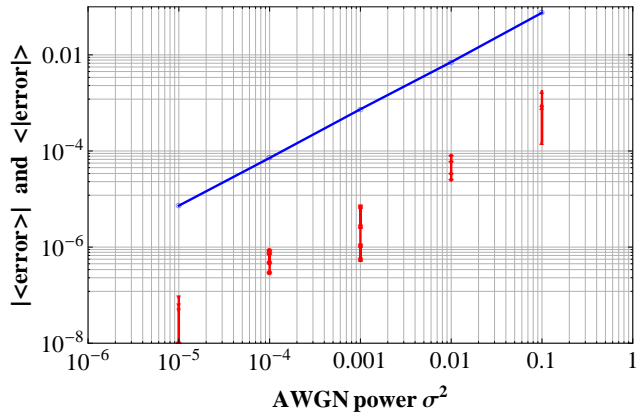}
\caption{(color online). Average PFFA error (dots) and average of the absolute value of the PFFA error (line); versus  AWGN noise power.}
\end{figure}

The utility of any parameter estimation algorithm will
lie in its ability to perform well under non-ideal conditions for the
collection of measurement statics. To that end we herein test the algorithm's resilience to thermal noise,
specifically: additive white-Gaussian noise (AWGN) in which five
statistically independent AWGN processes are added to each of the
the probabilities of the 5 different values of $m,$ $\{+2,+1,0,-1,-2\}$
for a  N00N state input of $j=2$  with $\Phi $ fixed. After each
addition of the five independent noise samples, the PFFA is executed and its
phase estimation error is calculated. Each dot in Figure 4 represents the
average error obtained from 40,000 such executions; and four such dots are
presented for each value of noise power $\sigma^{2}$ to demonstrate the
variation in the average error. The estimate is unbiased so there is some cancellation due to polarity in that average, which is useful in some applications. In other applications the absolute value of the  error is more relevant and these are connected by the line in Figure 4 (averaged over only 2,000 executions) and these did not vary appreciably from run to run. The line continues to drop linearly as the AWGN power is diminished (to a limit set only by classical effects: e.g., the choice of LMS versus other fits, the number of digits retained in the numerical processing, etc.). These results are applicable in the static limit. In the problem of tracking a dynamic phase shift the rapidity with which the histograms converge to their PDFs, and the quantum state dependent aspects of these (HWHM, V, etc.) will be of great significance.   \vspace{-.2in}

\section{Concluding remarks}
\vspace{- 0.1in}

Within the quantum phase representation we have derived Heisenberg limits in closed form for three classes of states in terms of two local performance metrics: HWHM and bin-variance. One of these three comprises the N00N states and we demonstrated how sub-states can employ sub-harmonics to outperform these in both metrics. The sub-states also increase the peaks of retained bins (a property which could be useful for tracking a dynamic phase) while diminishing alternate bins which can be dropped with a minimal probability of a dropped estimate equal to less than 3.164\%. All three classes of states are Heisenberg limited in that the HWHM and the square-root of the bin-variance scale as $1/N$ (where $N$ is the expected number of photons) but the coefficients multiplying $1/N$ can vary. The third class of states (termed N00N-vac states) entangle the vacuum state with a N00N state, of relative probability equal to n. For fixed $N$, these enhance the super-resolution (number of fringes/bins) by a factor of $(n+1)/2$ and for large $n$: they diminish the bin-variance as $1/n^{2}$ relative to the N00N state results at a diminished fringe visibility which scales as $1/\sqrt{n}$. These higher-order (super-resolving) fringes vanish from the output signal of a first-order interferometer, in which one averages the difference of the two photodetector currents, so coincidence detection schemes of order $N$ have been utilized. We discussed an algorithm which can extract this higher-order information from the apparatus of a standard interferometer (of complexity independent of $N$) by processing the information in a way that incorporates knowledge of the allowable quantum results. The algorithm provides over nine digits of phase estimation accuracy from a four-photon N00N state (over an unknown signal range of $\pi/4$) within a standard interferometer; and is shown herein to be fairly robust to the presence of additive white-Gaussian noise in the measurement statistics. 

\vspace{.15in}
\hspace{-0.13in}{\textbf{Acknowledgments}}
\vspace{.15in}

\noindent S.R.S. would like to acknowledge support from LED.
F.I.M. would like to acknowledge support from NASA EPSCoR \&
LaSPACE, Louisiana. J.P.D. would like to acknowledge support from the Air Force Office of Scientific Research and the Army Research Office.


\begin{thebibliography}{99}

\bibitem{1} P.W. Shor, 35th Annual Symposium on Foundations of Computer Science, Santa Fe (1994).

\bibitem{2} M. Wilde, Quantum Information Theory, Cambridge University Press (2013).

\bibitem{3} C. G. Timpson, Quantum Information Theory and the Foundations of Quantum Mechanics, Oxford University Press (2013).

\bibitem{4} R. Cleve, A. Ekert, C. Macchiavello, M. Mosca, Proc. R. Soc. Lond. A \textbf{454}, 339 (1998).

\bibitem{5} A. Ekert, Physica Scripta \textbf{76}, 218 (1998).

\bibitem{6}R.  Feynman,  Int. J. Theor. Phys. \textbf{21}, 467 (1982). 

\bibitem{7} H. Lee, P. Kok, J.P. Dowling, J. Mod. Optic. \textbf{49}, 2325 (2002).

\bibitem{8} E. Knill, R. Laflamme, G.J. Milburn, Nature \textbf{409}, 46 (2001).

\bibitem{9} M. J. Holland and K. Burnett, Phys. Rev. Lett. \textbf{71}, 1355 (1993).

\bibitem{10} B. C. Sanders and G. J. Milburn, Phys. Rev. Lett. \textbf{75}, 2944 (1995).

\bibitem{11}  J. P. Dowling, Contemporary Physics \textbf{49},125 (2008).

\bibitem{12} A. N. Boto, P. Kok, D. S. Abrams, S. L. Braunstein, C. P. Williams, J. P. Dowling, Phys. Rev. Lett. \textbf{85} , 2733 (2000).

\bibitem{13}  S. Guha and J. H. Shapiro, Quantum Communication, Measurement and Computing, AIP Conf. Proc. \textbf{1363}, 113 (2011).

\bibitem{14} M. D'Angelo, M. V. Chekhova, and Y. Shih, Phys. Rev. Lett. \textbf{8701}, 013602 (2001).

\bibitem{15} M. W. Mitchell, J. S. Lundeen, and A. M. Steinberg, Nature \textbf{13}, 161 (2004).

\bibitem{16} P. Walther, J. W. Pan, M. Aspelmeyer, R. Ursin, S. Gasparoni and A. Zeilinger, Nature \textbf{429}, 158 (2004).

\bibitem{17} By ``standard interferometer'' we refer to the apparatus of a ``first-order interferometer'' in which the two photodetector currents are not necessarily subtracted. If they are subtracted then it is a ``first-order interferometer;'' if they are multiplied then it is a ``second-order interferometer.'' If it is a higher-order interferometer then more beam splitters and more photodetectors are required in the apparatus than in the standard interferometer. 

\bibitem{18} S. R. Shepard, Nonlinear Analysis \textbf{71}, 1160  (2009).

\bibitem{19} S. R. Shepard, Ph.D. thesis, MIT, Cambridge, MA (1992).

\bibitem{20} S. R. Shepard, ``Fundamental Problems in Quantum Theory,'' Proceedings New York Academy of Sciences \textbf{755}, 812 (1995). 

\bibitem{21} B. Yurke, S. L. Mc Call and J. R. Klauder, Phys. Rev. A \textbf{33}, 4033 (1986).

\bibitem{22} J. Schwinger, U.S. Atomic Energy Report No. NYO-3071 U.S. GPO, Washington, D.C. (1952).

\bibitem{23} For $\Phi $ beyond a
bin-width the algorithm can and sometimes does make a bin error (converging
to what would be the right answer but displaced to the wrong branch of the
periodic PDF) if we do not give it any apriori knowledge to avoid these.

\bibitem{24} P. M. Anisimov, G. M. Raterman, A. Chiruvelli, W. N. Plick, S. D. Huver, H. Lee, J. P. Dowling, Phys. Rev. Lett. \textbf{104}, 103602 (2010).
\end{thebibliography}
\end{document}